\documentclass[aip,cha,english,reprint,a4paper]{revtex4-2}

\usepackage[T1]{fontenc}
\usepackage{graphicx}
\usepackage{dcolumn}
\usepackage{bm}

\usepackage{amsmath,amssymb}
\usepackage{mathptmx}
\usepackage{physics}
\usepackage{etoolbox}
\usepackage[caption = false]{subfig}
\usepackage{standalone}
\usepackage{xcolor}
\usepackage{enumitem}
\usepackage{mathtools}
\usepackage{makecell}
\usepackage[colorlinks = true,linkcolor = red,citecolor = magenta]{hyperref}
\usepackage[sort&compress]{natbib}

\newcommand{\romeadd}{Physics Department, Sapienza University of Rome, Piazzale Aldo Moro 5, Rome, 00185, Italy}

\newcommand{\rutadd}{Department of Physics and Astronomy, Center for Materials Theory, Rutgers University, Piscataway, NJ, 08854, USA}

\newcommand{\coladd}{Physics Department, Columbia University, New York, NY, 10027, USA}

\newcommand{\pcsadd}{Center for Theoretical Physics of Complex Systems, Institute for Basic Science (IBS), Daejeon, 34126, Republic of Korea}

\newcommand{\ustadd}{Basic Science Program, Korea University of Science and Technology (UST), Daejeon, 34113, Republic of Korea}

\makeatletter
\def\@email#1#2{
  \endgroup
    \patchcmd{\titleblock@produce}
      {\frontmatter@RRAPformat}
      {\frontmatter@RRAPformat{\produce@RRAP{*#1\href{mailto:#2}{#2}}}\frontmatter@RRAPformat}
      {}{}
}%
\makeatother

\newcommand{\beg}{\begin{equation}}
\newcommand{\en}{\end{equation}}

\newcommand{\eref}[1]{Eq.~(\ref{#1})}

\usepackage{amsmath}
\usepackage[bb=dsserif]{mathalpha}
\usepackage{bm}

\begin{document}

\preprint{AIP/123-QED}

\title[Dynamical chaos in the integrable Toda chain induced by time discretization]{Dynamical chaos in the integrable Toda chain induced by time discretization}

\author{Carlo Danieli}
  \email{danielic@roma1.infn.it}
  \affiliation{\romeadd}

\author{Emil A. Yuzbashyan}
  \affiliation{\rutadd}

\author{Boris L. Altshuler}
  \affiliation{\coladd}  

\author{Aniket Patra}
  \email{patraaniket@gmail.com}
  \affiliation{\pcsadd}
    
\author{Sergej Flach}
  \email{sflach@ibs.re.kr}
  \affiliation{\pcsadd}
  \affiliation{\ustadd}

\date{\today}

\begin{abstract}
We use the Toda chain model to demonstrate that numerical simulation of integrable Hamiltonian dynamics using time discretization destroys integrability and induces dynamical chaos. Specifically, we integrate this model with various symplectic integrators parametrized by the time step $\tau$ and measure the Lyapunov time $T_{\Lambda}$  (inverse of the largest Lyapunov exponent $\Lambda$).  A key observation is that $T_{\Lambda}$ is finite whenever $\tau$ is finite but diverges when $\tau \rightarrow 0$. 
We compare the Toda chain results with the nonitegrable Fermi-Pasta-Ulam-Tsingou chain dynamics. In addition, we observe a breakdown of the simulations at times $T_B \gg T_{\Lambda}$  due to certain positions and momenta becoming extremely large (``Not a Number''). This phenomenon originates from the     periodic driving introduced by  symplectic integrators and we also identify the concrete mechanism of the breakdown in the case of the Toda chain.
\end{abstract}

\maketitle

\begin{quotation}
Classical integrable systems have vanishing Lyapunov exponents.  State-of-the-art computational tests of the equations of motion employ symplectic integrators (SI). Such split-step integrators replace the original Hamiltonian by some time-dependent one which is parametrized by the finite integrator time step $\tau$. It follows that SIs in general replace integrable dynamics by a nonintegrable chaotic one. We analyze the manifestation of chaos using two different split-step symplectic integrators ($ABA864$ and $SABA_2$) for the integrable Toda chain with fixed ends. We demonstrate that the system is indeed chaotic on large times and has a positive maximum Lyapunov exponent $\Lambda$.
The extracted Lyapunov time $T_\Lambda = 1/\Lambda$ signals the onset of dynamical chaos, is $\tau$-dependent, and diverges for $\tau \rightarrow 0$. For small time-steps $\tau$, up to a much larger time  $T_E$, the energy fluctuations stay bounded (while other Toda integrals do not), which means that the SIs emulate a new nonintegrable Hamiltonian. For even larger times,  
we observe a Floquet regime, when the system exhibits unrestrained heating. This, in turn, unleashes rogue fluctuations leading to very large values of coordinates and momenta and the eventual breakdown of the numerical integration procedure at some time $T_B$. 
\end{quotation}





\section{Introduction}\label{sec:intro}

The computational study of many-body dynamics has been at a cornerstone of exploring the physics of interacting many-particle systems including gases,  liquids, and solids and attempting to understand   properties of the liquid-glass transition and of laminar and turbulent flows,  dynamics of defects,  etc.~\cite{1983rsm,Heermann1990Springer}. For classical problems, this often means introducing a Hamiltonian
$H({\bf q},{\bf p})$ with $N$ degrees of freedom, where  ${\bf q} = (q_1,\dots,q_N)$  are the canonical coordinates  and $ {\bf p} = (p_1,\dots,p_N)$ are the corresponding momenta, and solving a set of coupled Hamilton's equations of motion
\begin{equation}
\label{Hamilton}
\dot{{\bf q}}= \frac{\partial H}{\partial {\bf p}}\ ,\quad
\dot{{\bf p}}=- \frac{\partial H}{\partial {\bf q}},
\end{equation}
where the dot stands for the derivative with respect to  the continuous time variable $t$.

Since the dawn of computers with von-Neumann architecture, numerical integration of nonlinear differential equations is typically  performed by discretizing the time $t$ into short intervals $\tau$ and attempting to integrate the equations sequentially over one time interval after another   until a sufficiently large final integration time is reached~\cite{abr_steg64,Heermann1990Springer}. Thus, continuous temporal phase-space flows are replaced by a repeated action of discrete maps acting on the phase-space variables of the system. Examples are the Euler,  Runge-Kutta, and  Verlet
(also known as Leap-Frog~\cite{Heermann1990Springer}) algorithms. 
In general, the energy conservation law $dH/dt=0$, which follows straightforwardly from (\ref{Hamilton}), is violated by such maps. Symplectic maps (with the Verlet algorithm being an early example) preserve the phase-space volume and the Hamiltonian nature of the dynamics, albeit with a time-dependent effective Hamiltonian.  

Usually, a  Hamiltonian $H({\bf q},{\bf p})$ can be split into two parts, $H=A+B$, such that the exact dynamics of $A$ and $B$ alone is known analytically and explicitly. For example, this is the case when $A$ depends only on the momenta (kinetic energy), while $B$ is a function of the coordinates only (potential energy). Even though the dynamics of $A$ and $B$ are separately exactly solvable,  solving for the dynamics of the full Hamiltonian $A+B$ is still a nontrivial problem as long as the Poisson bracket
\begin{equation}
\label{Poisson}
\{A,B\}= \frac{\partial A}{\partial {\bf q}} \frac{\partial B}{\partial {\bf p}} - \frac{\partial B}{\partial {\bf q}} \frac{\partial A}{\partial {\bf p}}
\end{equation}
does not vanish. Indeed, it is well know that a typical Hamiltonian with $N>1$ degrees of freedom is nonintegrable. At the same time, there are notable examples of  nontrivial integrable systems, whose dynamics is characterized by $N$ conserved quantities (integrals of motion). 

Numerical integration of the Hamiltonian dynamics is usually performed using split-step methods, which  break the   time $t$ into  short time intervals of lengths proportional to $\tau$, e.g., intervals of lengths $\frac{\tau}{2},\tau,\frac{\tau}{2},\frac{\tau}{2},\tau,\frac{\tau}{2}\dots$ or simply $\tau,\tau,\tau\dots$    and time   evolve the system  with $A$ and $B$ intermittently, i.e., evolve    with $A$ over some of the time intervals and   with $B$ over the rest. Note that  as $\tau\to0$, the so discretized dynamics converges to the continuous time dynamics with the original Hamiltonian $H$.  Below we focus on such symplectic split-step integration schemes. Details about implementing such symplectic integrators can be found, e.g., in Refs.\ \onlinecite{Yoshida1990construction, danieli2019computational}. 

The errors and deviations due to discretizing  Hamiltonian dynamics  appear to be well analyzed and estimated \cite{Yoshida1990construction}. However, note that since time discretization    replaces the time-independent Hamiltonian $H$ with a  time-dependent (periodic) one, it changes the degree of chaoticity of the system  as measured, e.g., by the largest Lyapunov exponent $\Lambda$. The impact  of this is likely to be most dramatic when the original Hamiltonian is integrable.  In this case,  we expect to observe   \textit{emergent chaoticity}, because time discretization breaks the integrability by violating not only the energy conservation, but also the remaining $N-1$ conservation laws. Moreover, since the original Hamiltonian is integrable and its dynamics is therefore nonchaotic (regular),  the chaoticity  must emerge in an integrable system as a result of the discretization.  We therefore anticipate a nonzero $\tau$-dependent Lyapunov exponent $\Lambda(\tau)$, such that $\Lambda(\tau)\to0$ when $\tau\to0$. In addition, the energy is no longer conserved  since the actual Hamiltonian  is time dependent. As a result, there is no Gibbs distribution to protect the dynamics from rogue  fluctuations.

A number of publications focused on the impact of numerical integration schemes on the dynamics of integrable sine-Gordon models, nonlinear Schr\"{o}dinger equations, and also the spatially discrete Ablowitz-Ladick chain (an integrable discrete version of the space-continuous nonlinear Schr\"{o}dinger equation \cite{AblowitzHerbstSchoberPA1996,CaliniErcolaniMcLaughlinSchoberPD1996,AblowitzHErbstSchoberJCP1997,AblowitzOhtaTrubatchCSF2000,AblowitzHerbstSchoberJPA2001,IslasKarpeevSchoberJCP2001,SungMoonKimCPL2001,TriadisBroadbridgeKajiwaraMarunoSAM2018}). Most  efforts were directed towards  verifying  that time discretization does destroy integrability and induce the so-called numerical chaos without, however, a systematic quantification of this process through the computation of Lyapunov exponents. Further attempts were directed at finding better integrators which diminished the impact of numerical chaos (we thank R. McLachlan for pointing out that the ideal situation would be to discretize time working in the action-angle coordinate frame of the integrable system, which is however quite often a formidable task). 

In this work, we analyse the dynamics of the integrable Toda chain. It is one of the very few examples of one dimensional integrable nonlinear lattices. This model is used, for instance, to understand heat conduction of solids \cite{TodaHeatCond, TodaImpHeatCond}, thermally generated soliton dynamics in DNA \cite{DNADyn}, and the soliton dynamics on a hydrogen-bond network in helical proteins \cite{HBondProtein}. We perform a quantitative analysis of the impact of time discretization on the largest Lyapunov exponent. We find that its inverse -- the Lyapunov time -- is diverging power-law-like upon decreasing the step size. We also identify a second, much larger time scale $T_B$ -- the breakdown time --  at which a large fluctuation leads to the breakdown of the numerical simulation. We employ different symplectic integrators and show that the results are qualitatively independent of the integrator choice.

 We first introduce symplectic integrators in Sec.\ \ref{sec:symp}. We define the integrable Toda chain as well as one of its famous nonintegrable approximations -- the $\alpha$-Fermi-Pasta-Ulam-Tsingou (FPUT) chain in Sec.\ \ref{sec:toda}. In this section, we apply the ABA864 integrators to both models with a suitably  small step size and a finite observational time window.   This  confirms the usual result that the FPUT chain generates a measurable nonzero Lyapunov exponent. The Toda chain, on the other hand, appears to get along with a vanishing Lyapunov exponent as its finite time average decreases as $1/T$ with the integration time $T$. However, in Sec.\ \ref{sec:chaos}, we show that one obtains a finite Lyapunov exponent for the Toda chain as well if one considers large enough integration time $T$. In this section, we measure its dependence on the step size and then repeat the computation with a simpler SABA$_2$ integrator and observe a similar outcome. We also measure the breakdown time  at which fluctuations make further computation impossible as a function of the time step size. Finally, we conclude with a discussion of our results.

\section{Time discretization and symplectic integration of lattice Hamiltonians}\label{sec:symp}

We consider a set of $2N$ coupled first order differential equations 
generated by the Hamiltonian $H({\bf q},{\bf p})$ with $N$ degrees of freedom written for the canonical coordinates ${\bf q} = (q_1,\dots,q_N)$ and momenta $ {\bf p} = (p_1,\dots,p_N)$. 
The dynamics is given by the $2N$ Hamilton's equations of motion
\begin{equation}
\label{eq:ham_eq}
 \dot{{\bf z}} = L_H{\bf z} = \{H,{\bf z}\} 
 \qquad \Leftrightarrow \qquad 
\dot{{\bf q}}= \frac{\partial H}{\partial {\bf p}}\ ,\quad
\dot{{\bf p}}=- \frac{\partial H}{\partial {\bf q}}
\end{equation}
for ${\bf z} = ({\bf q},{\bf p})$, and the Liouvillian operator $L_H$ defined via the Poisson brackets 
(\ref{Poisson}).
It follows that
\begin{equation}
\label{eq:ham_eq_sol}
{\bf z}(t) = e^{t L_H}{\bf z}(0)
= \sum_{s=0}^{+\infty} \frac{t^s}{s!} \left(L_H\right)^s {\bf z}(0) .
\end{equation}
We use symplectic integration schemes for Eqs.~\eqref{eq:ham_eq} to approximate $e^{t L_H}$ in Eq.~\eqref{eq:ham_eq_sol} such that
the phase space volume is preserved.   
We follow Ref.~\onlinecite{laskar2001high} and consider a Hamiltonian $H$ which can be written as the sum of two parts $H= A + B$ such that
$e^{tL_A}$ and $e^{tL_B}$ are explicitly known in closed form.
The Baker-Campbell-Hausdorff formula approximates the operator $e^{\tau L_H}$,
\begin{equation}
e^{\tau L_H} = \prod_{j=1}^k  e^{a_j \tau L_{A}}e^{b_j \tau L_{B}}  + \mathcal{O}(\tau^{p}),
\label{eq:symp}
\end{equation}
where the
coefficients $a_1,b_1,a_2,b_2,...,a_k,b_k$ must satisfy $\sum_{j=1}^k a_j = \sum_{j=1}^k b_j = 1$. The accuracy of the approximation is controlled by the exponent $p$ which depends on $k$ and the  values of the coefficients.
 The simplest possible choices in Eq.~\eqref{eq:symp} are
  $k=1$ and  $a_1=b_1=1$, which gives $p=2$ ~\cite{Yoshida1990construction}. This integrator violates time reversal symmetry.  
The celebrated   Verlet  (a.k.a  Leap-Frog) scheme preserves the time reversal symmetry and is given by 
 $k=2$ with $a_1=a_2=\frac{1}{2}$, $b_1=1, b_2=0$, and $p=3$
 in Eq.~\eqref{eq:symp}: 
$e^{\tau L_H} = e^{\frac{1}{2}\tau L_{A}}e^{\tau L_{B}}e^{\frac{1}{2}\tau L_{A}} + \mathcal{O}(\tau^{3})$ ~\cite{Yoshida1990construction}.
Note that the dynamics described by Eq.~\eqref{eq:symp} corresponds to a periodic in time Hamiltonian. For example, in the simplest case  $k=1$ and  $a_1=b_1=1$, we have
\begin{equation}
H_\mathrm{eff}(t)= A+ (B-A) f(t),
\label{actual}
\end{equation}
where $f(t)$ is a periodic function of time equal to 0 for odd numbered time intervals and  1 otherwise, i.e.,
\begin{equation}
f(t)=
\begin{cases}
0\quad \mbox{ for $2k\tau < t < (2k+1)\tau,$}\\
1\quad \mbox{ for $(2k+1)\tau < t < (2k+2)\tau,$}\quad k=0,1,2\dots\\
\end{cases}
\end{equation}

Suppose the original Hamiltonian $H$ is integrable and therefore possesses $N$ nontrivial   integrals of motion. 
The   Hamiltonian $H_\mathrm{eff}(t)$ in Eq.~\eqref{actual} with which we actually evolve the system is most likely   nonintegrable  and violates all  of the above $N$ conservation laws for any $\tau\ne0$, including the energy conservation as it is time dependent. We thus expect the dynamics to become  chaotic and 
 characterized by nonzero Lyapunov exponents. 
 The time step $\tau$ acts as an effective strength of the  integrability breaking perturbation. 
 In what follows, we will test these ideas on the
classical integrable   Toda  chain and evaluate the largest Lyapunov exponent as a function of the time-step $\tau$ for two different symplectic schemes.

\section{The Toda Chain}\label{sec:toda}

\begin{figure}
    \centering
    \includegraphics[scale=0.45]{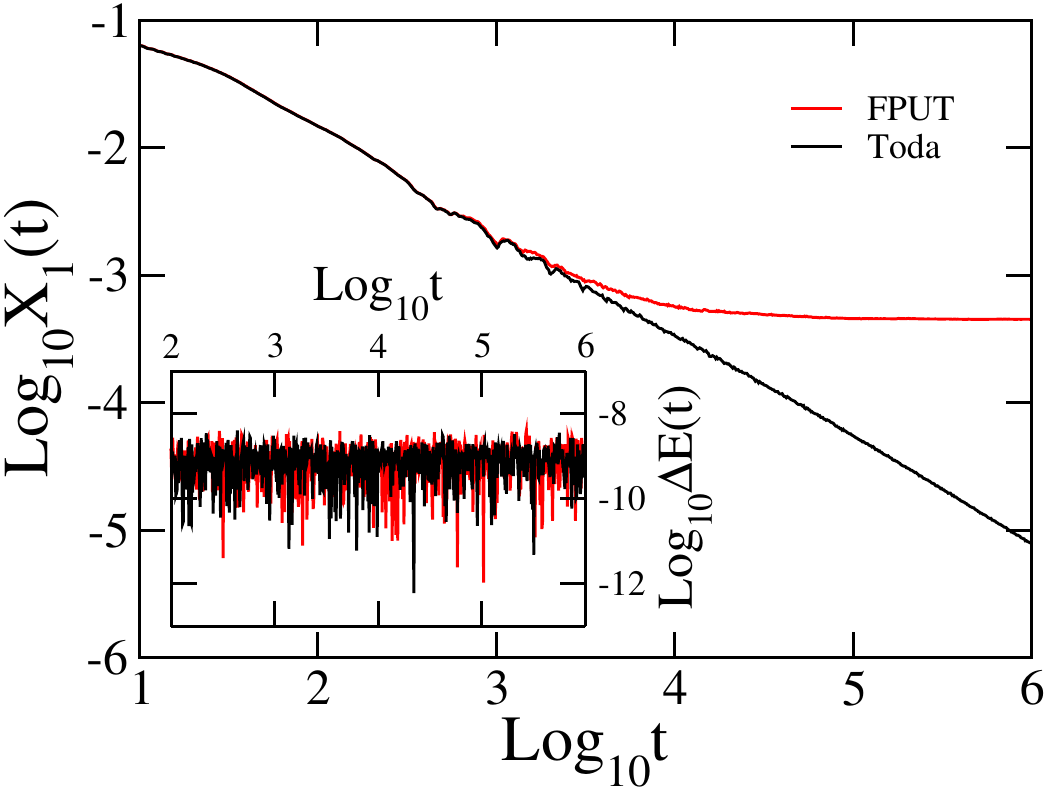}
    \caption{Time evolution of the finite-time largest Lyapunov exponent $X_1(t)$  for the Toda  (black) and FPUT (red) chains  obtained with the symplectic  integrator $ABA864(\tau )$ with time-step $\tau=0.1$, $N=100$ particles, and the anharmonicity strength $\alpha=0.25$.  
    Inset: relative energy error $\Delta E(t)$. 
    }
    \label{fig:1}
\end{figure} 

The Toda chain is an integrable one-dimensional lattice model defined by the
Hamiltonian~\cite{Toda1975studies}
\begin{equation}
\begin{split} 
\label{eq:sep_ham}
& H_\mathrm{T} = \sum_{n=1}^{N} \left[ \frac{p_n^2}{2} + V_\mathrm{T}(q_{n+1}-q_n)\right],\\
& V_\mathrm{T}(r) = \frac{1}{4\alpha^2}(e^{2\alpha r}-2\alpha r-1).
\end{split}     
\end{equation}
 Its 
 integrability  for  fixed and for periodic boundary conditions  was proved in Refs.~\onlinecite{Ford1973integrability,Henon1974integrals,Flaschka1974toda}. The parameter $\alpha$ can be absorbed by a rescaling of coordinates and momenta. We prefer to keep it in order to make it easier for the reader to connect to data of previous studies.


In this section, we will follow the standard approach of comparing the dynamics of the integrable Toda chain with fixed ends (fixed boundary conditions) and the
 Fermi-Pasta-Ulam-Tsingou model (FPUT). We do so  by computing the corresponding largest Lyapunov exponents. The Hamiltonian $H_\mathrm{FPUT}$ of the FPUT model is a low energy approximation of the Toda chain obtained  by replacing $V_\mathrm{T}(r)$ in Eq.~\eqref{eq:sep_ham} with its  truncated Taylor expansion 
\begin{equation}
V_\mathrm{FPUT}(r) = \frac{1}{2} r^2 + \frac{\alpha}{3} r^3,
\label{FPUT}
\end{equation}
see, e.g., Refs.~\onlinecite{PonnoChristodoulidiSkokos,Benettin2018FPUT}. This innocent approximation destroys integrability.
 The FPUT chain was used for the first computational studies of thermalization \cite{FPUT-CP}, lead to the discovery and  naming of solitons~\cite{Zabusky1965interaction,Zabusky1967dynamics}, 
was continuously used in subsequent studies of thermalization and equipartition~\cite{Casetti1997FPUT,Benettin2011time,Danieli2017intermittent,Lvov2018double}, and served as a platform for a plethora of other studies,
for reviews see Refs.~\onlinecite{Ford1992FPUT,Porter2009birth}.

We apply symplectic integration schemes 
since both the kinetic part $A$ and the potential part $B$ of $H_\mathrm{T}$ and $H_\mathrm{FPUT}$ are integrable\footnote{The kinetic and potential parts $A$ and $B$ in Eq.~\eqref{eq:sep_ham} depend only on the momenta ${\bf p}$ and positions ${\bf q}$, respectively. Hence, $A$ conserves each of the $N$ momentum coordinates $p_n$, while $B$ conserves each of the $N$ position coordinates $q_n$.}. 
We report the explicit form of the resolvent operators $e^{\tau L_{A}}$ and $e^{\tau L_{B}}$ in Eq.~\eqref{eq:symp} for both  Toda and  FPUT chains in  Appendix~\ref{sec:A1}. 
For a symplectic integrator, we choose  a fourth-order, $p=4$, scheme called $ABA864$ and introduced in Ref.~\onlinecite{BLANES201358} (its explicit form and the coefficients $a_j,b_j$ are reported in  Appendix~\ref{sec:A2a}). 
This integrator has been highlighted in Ref.~\onlinecite{danieli2019computational} as one of the best performing in terms of accuracy, stability, and efficiency among several other symplectic and non-symplectic methods. 
As commonly done when computing the propagation of Hamiltonian systems, we check the stability and accuracy of the symplectic integrator by evaluating the relative error in the Hamiltonian energy $H$
\begin{equation}
\label{eq:rel_err}
\Delta E(t) = 
\bigg\lvert \frac{H(t) - H(0)}{H(0)} \bigg\rvert.
\end{equation} 
A simulation is considered accurate and stable if this quantity $\Delta E$, while fluctuating, remains below a set precision threshold $E_r$ of the order of the first nonzero value of Eq.~\eqref{eq:rel_err} over the whole integration time-window $[t_0,t_0+T]$. In the field of classical lattice dynamics, $E_r=10^{-5}$ is typically considered  an upper bound for a good accuracy threshold. In contrast, the  consistent growth of $\Delta E$ and subsequent  breach of the threshold  $E_r$ indicate the loss of accuracy of a method\footnote{This type of a test on the relative error may be extended to other conserved quantities, such as the total norm and the total momentum.}.

We compute the largest Lyapunov exponent $\Lambda$ and its inverse,   a.k.a. Lyapunov time $T_\Lambda=\frac{1}{\Lambda}$ by numerically integrating the variational equations associated with the Hamilton's equations~\eqref{eq:ham_eq}. 
As explained in detail in Ref.~\onlinecite{Skokos2010numerical}, this means  computing the time evolution of a small deviation vector ${\bf w}(t)= (\delta {\bf q}(t),
\delta {\bf p}(t))=\left(\delta q_1 (t), \ldots,
  \delta q_N (t), \delta p_1(t), \ldots, \delta p_N (t)
\right) = (x_i)_{i=1}^{2N}$ for a trajectory ${\bf z}(t)$.
The variational equations describing the evolution of ${\bf w}(t)$ read
\begin{equation}
\label{eq:var_eq}
\dot{{\bf w}}(t) =\big[ {\bf J}_{2N} \cdot {\bf D}_{ H}^2 ({\bf z}(t) ) \big] \cdot {\bf w} (t).
\end{equation}
Here ${\bf D}_{ H}^2 ({\bf z}(t) )$ is the Hessian matrix of the Hamiltonian  $H$
computed on the phase-space trajectory ${\bf z}(t)$
\begin{equation}
\label{eq:hessian}
\left[{\bf D}_{H}^2 ({\bf z}(t)) \right]_{i, j} =
\frac{\partial^2 H}{\partial x_i \partial x_j}\bigg\vert_{{\bf z}(t)}\, \quad i,j=1,\ldots,2N,
\end{equation}
${\bf J}_{2N}$ is the symplectic identity
\begin{equation}
\label{eq:symp_max}
{\bf J}_{2N} =
\begin{bmatrix}
    \mathbb{O}_N  &  \mathbb{1}_N \\
   - \mathbb{1}_N  &  \mathbb{O}_N
\end{bmatrix},
\end{equation}
and $\mathbb{1}_N$ and $\mathbb{O}_N$ are the identity and  null matrices of rank $N$, respectively. 
Eq.~\eqref{eq:var_eq} can be explicitly solved for a split-step time evolution like ours,  
and we report the variational problems and the corresponding resolvents for  Toda and FPUT models in  Appendix~\ref{sec:A1}. 

The largest Lyapunov exponent $\Lambda$ is obtained by first computing the so-called {\it finite-time largest Lyapunov exponent} defined as
\begin{equation}
X_1(t) = \frac{1}{t}\ln  \frac{\|{\bf w}( t)\|}{\|{\bf w}(0)\|},
\label{eq:finitetimeLE}
\end{equation}
and then taking the limit $t\rightarrow +\infty$,
\begin{equation}
\Lambda= \lim_{t\rightarrow +\infty} X_1(t).
\label{eq:lambda}    
\end{equation}
The {\it largest Lyapunov exponent} $\Lambda$ and the corresponding {\it Lyapunov time} $T_\Lambda = \frac{1}{\Lambda}$ set the timescale on which a dynamical system becomes chaotic. 

In Fig.~\ref{fig:1} we plot the time evolution of the finite-time largest Lyapunov exponent $X_1(t)$  for both Toda and FPUT models for the same initial state with energy density $\varepsilon = \frac{H}{N} = 0.1$\footnote{\label{ftnt1}We obtain the initial state   through the following procedure. First, set $q_n=0$ and select $p_n$  randomly  from a uniform distribution in the interval $[-1,1]$. Next, uniformly  rescale all $p_n$ so that the energy density $\varepsilon=0.1$ and eviolve this configuration  for $T=10^5$ time steps with $ABA864(\tau=0.01 )$. The resulting $\{q_n,p_n\}$  is the initial state for the simulations.}, 
 fixed integration step $\tau=0.1$, $N=100$ and $\alpha=0.25$. 
The   $X_1(t)$ curves  show a clear distinction between the Toda  (integrable, black color) and  FPUT (nonintegrable, red color) cases. Indeed, for  Toda  we observe  $X_1(t)\sim \frac{1}{t}$ within the observation time window. This seems to suggest   that the largest Lyapunov exponent vanishes, $\Lambda = \lim_{t\rightarrow \infty} X_1(t)=0$, as expected for an integrable system. In contrast, for the FPUT  chain $X_1(t)$ saturates at a finite   value resulting in a   finite largest Lyapunov exponent $\Lambda\approx 4.75\cdot10^{-4}$, which corresponds to a Lyapunov time $T_\Lambda\approx 2\cdot 10^3$.
In the inset in Fig.~\ref{fig:1}, we show that the relative energy error $\Delta E(t)$ obtained with the integrator $ABA864$ for a time-step $\tau=0.1$ oscillates  well below the precision threshold $E_r=5\cdot 10^{-9}$ for both systems. 

In what follows, we demonstrate that the Lyapunov time for the time-discretized Toda dynamics is in fact finite, and the largest Lyapunov exponent is nonzero for a finite step size $\tau$ and enlarged computational time windows.

\section{Time discretization induced chaos}\label{sec:chaos}

\begin{figure*}
    \centering
    \includegraphics[scale=0.75]{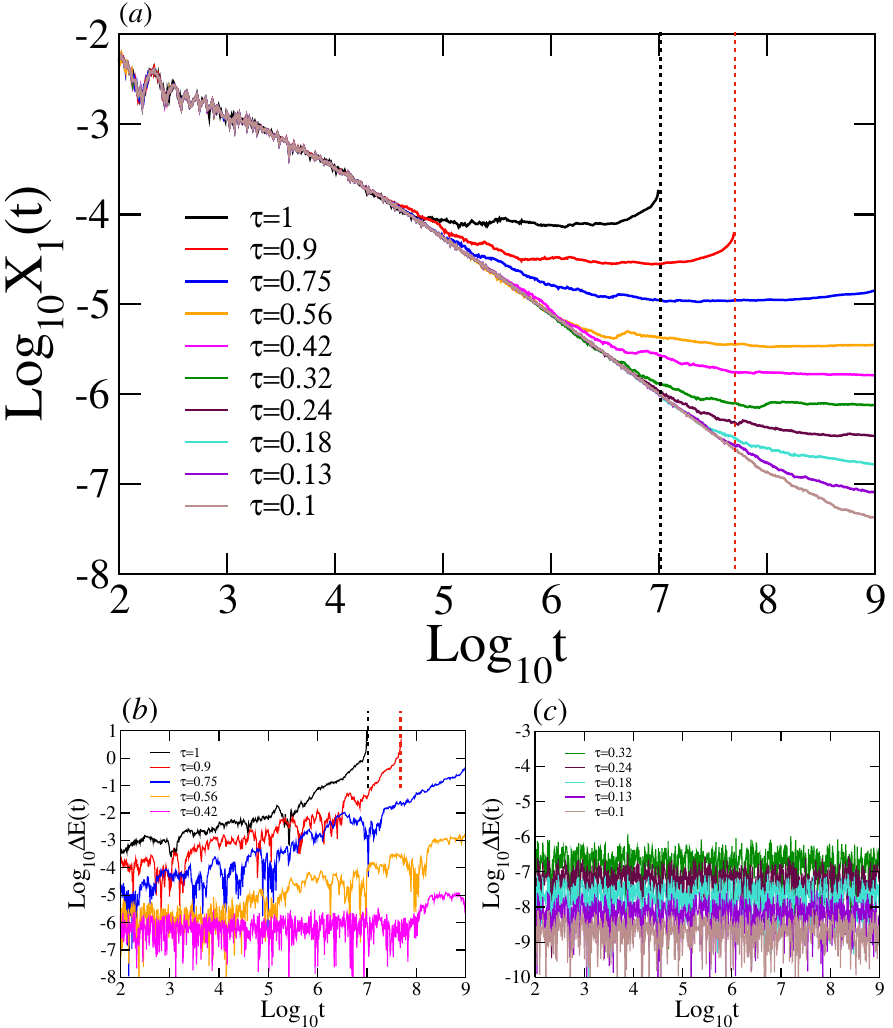}
    \caption{Chaoticity of the time-discretized  dynamics of the Toda chain with fixed ends generated by the symplectic  integrator $ABA864$. We show the time evolution of the  finite-time largest Lyapunov exponent $X_1(t)$ (a) and relative energy error $\Delta E(t)$ (b-c)  for $N=100$ sites and anharmonicity strength $\alpha=0.25$. We employed   the $ABA864$  integrator with time-step $\tau=1.0$ (black), $\tau=0.9$ (red), $\tau=0.75$ (blue), $\tau=0.56$ (orange), $\tau=0.42$ (magenta), $\tau=0.32$ (green), $\tau=0.24$ (maroon), $\tau=0.18$ (turquoise), $\tau=0.13$ (violet), and $\tau=0.1$ (brown).  
    The vertical dashed lines in (a,b) indicate the breakdown time $T_B$. 
    }
    \label{fig:2}
\end{figure*}

The results shown in Fig.~\ref{fig:1} align with the theory expectations as the two models produce clearly distinguishable dynamics. The nonintegrable FPUT case displays a nonzero largest Lyapunov exponent $\Lambda\approx 4.75 \cdot 10^{-4}$, while for the integrable Toda model  $X_1(t)\sim \frac{1}{t}$  apparently suggesting  that $\Lambda=0$. 
Furthermore, in both computations the integration is accurate, since the relative energy error $\Delta E$ stays well below $E_r=5\cdot 10^{-9}$. 
However, as conjectured in the Introduction, nonzero Lyapunov exponents are expected for large enough integration time $T$ when computing the dynamics of integrable systems, and the magnitudes of these exponents may vary  with the time-step $\tau$.
To demystify this conjecture, we extend the simulations for the Toda chain reported in Fig.~\ref{fig:1}  to a larger time window and different time steps $\tau$.

We computed the time evolution of the finite-time largest Lyapunov exponent $X_1$ within a time window $[0,10^9]$ for ten different time-steps $\tau$ ranging between $\tau=1$ and $\tau=0.1$ for   $N=100$ and $\alpha=0.25$ as in Fig.~\ref{fig:1}. We display the results 
in Fig.~\ref{fig:2}(a). Observe that within this time window   $X_1(t)$   clearly deviates from the naively expected $X_1(t)\sim\frac{1}{t}$ behavior for all $\tau$ considered. 
This deviation occurs earlier (and consequently $X_1(t)$ converges to a larger nonzero value)  for larger time steps $\tau$.

For $0.42 \leq \tau \leq 1$, the  relative energy error $\Delta E(t)$ starts to grow visibly from some time $T_E$ [Fig.~\ref{fig:2}(b)]. This time $T_E$ is different for each curve, which is characterized by having different values of $\tau$. For $0.1 \leq \tau \leq 0.32$ on the other hand, $\Delta E(t)$  in the entire time window fluctuates  below  the threshold $E_r$ that ranges from $E_r=10^{-6}$ for $\tau=0.32$ to $E_r=5\cdot 10^{-9}$ for $\tau=0.1$, i.e. $T_E > 10^9$ [see Fig.~\ref{fig:2}(c)].  
The latter case ($0.1 \leq \tau \leq 0.32$)   shows that the transition to chaotic dynamics (as detected by the maximum Lyapunov exponent $\Lambda$) occurs on a time scale much shorter than the time scale at which any noticeable increase of the relative energy error $\Delta E(t)$ is detected. 
The former case ($0.56 \leq \tau \leq 1$) demonstrates not only  the loss of accuracy of the symplectic scheme, but also a complete failure of the numerical integration for $\tau=1$ (black) and  $\tau=0.9$ (red). This failure takes place at a distinct breakdown time $T_B \gg T_{\Lambda}$.
 
 Failures of this type occur as certain  coordinates or momenta within the chain increase beyond $\sim 10^{308}$ and become NaN (Not-a-Number) at $T_B$.  To understand the nature of this effect, recall that the discretized time evolution with split-step symplectic integrators is governed by a time-periodic Hamiltonian, see Eq.~\eqref{actual}. The breakdown phenomenon originates from the fact that periodically driven many-body Floquet systems heat up indefinitely in the absence of disorder eventually reaching a featureless maximum entropy (infinite temperature) state~\cite{PhysRevX.4.041048,PhysRevE.90.012110,PONTE2015196,10.21468/SciPostPhys.3.4.029}. 
 If that ergodic system lacks any other conservation laws (note that energy conservation is already violated), it will explore the entire phase space, which is dominated by extremely large values of the coordinates and momenta\footnote{Note that   true Toda dynamics will not break down in this way, because  energy conservation prevents extremely large values of coordinates and momenta  in this case. On the other hand, for models such as the FPUT chain, where the potential is not bounded from below the energy conservation does not offer such a protection~\cite{CaratiPonno}. }. Consequently, we expect that   eventually  at least one of these quantities will diverge to infinity, leading to a failure of the integration on the computer (which can hold only floating point numbers up to some largest software dependent  value, e.g., $\sim 10^{308}$ in double precision with standard Fortran compilers). Later in this section we will identify the precise mechanism of the breakdown specifically for the time-discretized Toda dynamics.

To better understand all the time scales involved, we start from the shortest, which is $T_\Lambda$. For $t < T_\Lambda$ the dynamics remains integrable. The largest time scale is $T_B$, beyond which the computation ceases to be meaningful (see more below). Floquet dynamics results in an intermediate time scale $T_E$ at which Floquet heating and energy growth start.
 Necessarily $T_\Lambda < T_E < T_B$. By definition symplectic integrators are not constructed to approximately preserve any other integrals of motion other than the energy. Therefore another Toda integral $J$ (see Appendix~\ref{sec:A4}) will be bounded in their fluctuations only up to a time $T_J \approx T_\Lambda$. We discuss this issue in detail below Eq.\ \eqref{eq:toda_eq_dim_less}.

\begin{figure}
    \centering
    \includegraphics[width=0.85\columnwidth]{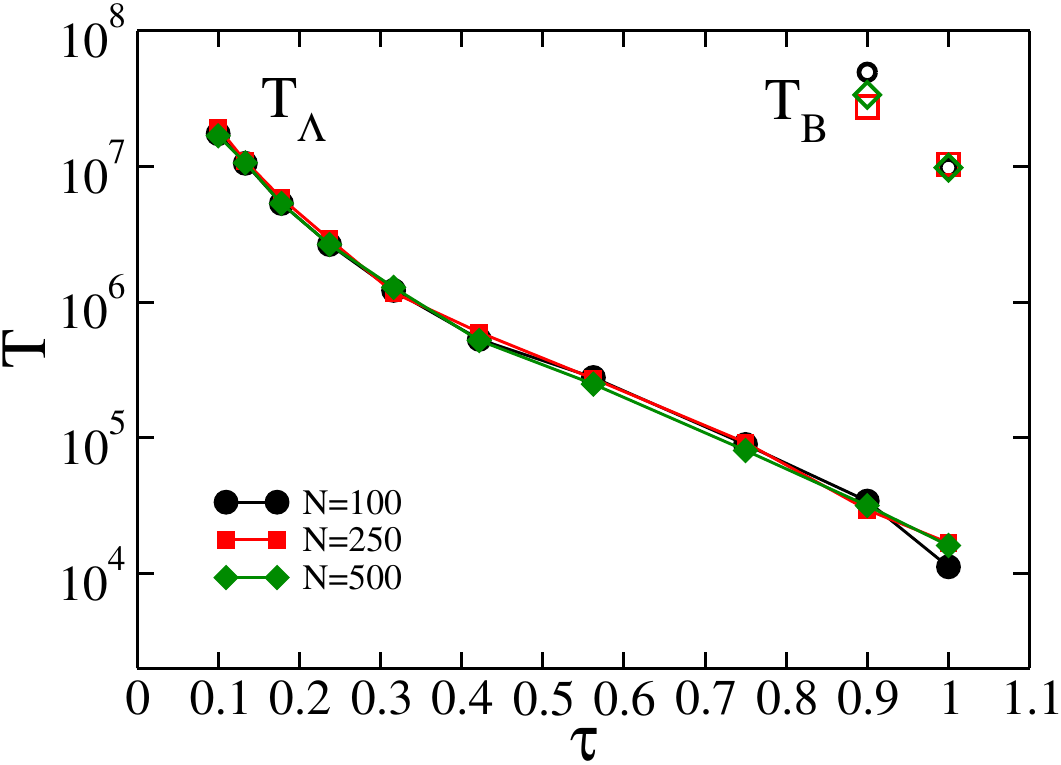}    
    \caption{Lyapunov time $T_\Lambda$ (solid symbols) and breakdown time $T_B$ (open symbols) versus the time-step $\tau$ for the   Toda chain dynamics as generated by the symplectic  integrator $ABA864(\tau )$ for $N=100$ (black dots), $N=250$ (red squares),  and $N=500$ (green diamonds) sites and anharmonicity parameter $\alpha=0.25$. 
    }
    \label{fig:3}
\end{figure}

\begin{figure*}
    \centering
    \includegraphics[scale=1.15]{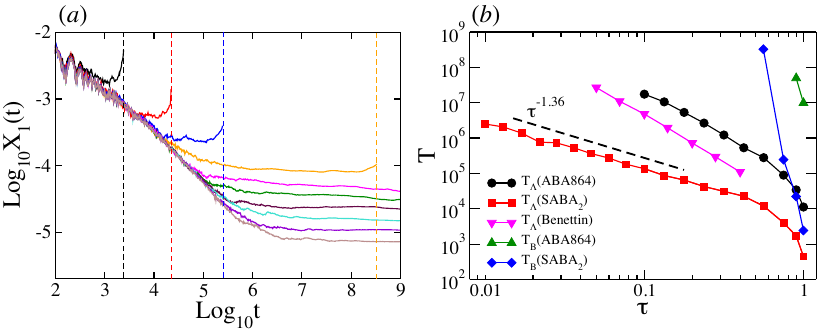}
    \caption{ Discretized time evolution of the Toda chain with $N=100$ sites and anharmonicity parameter $\alpha=0.25$. Panel  (a):  the finite-time largest Lyapunov exponent $X_1(t)$  obtained with the help of the symplectic integrator $SABA_2$  with time step $\tau=1.0$ (black), $\tau=0.9$ (red), $\tau=0.75$ (blue), $\tau=0.56$ (orange), $\tau=0.42$ (magenta), $\tau=0.32$ (green), $\tau=0.24$ (maroon), $\tau=0.18$ (turquoise), $\tau=0.13$ (violet), and $\tau=0.1$ (brown). 
  Panel  (b): Lyapunov times $T_\Lambda$ for integrators $ABA864$ (black circles) and $SABA_2$ (red squares) and breakdown times $T_B$ for $ABA864$ (green up triangles) and $SABA_2$ (blue diamonds)  versus the time-step $\tau$. 
    Magenta down triangles indicate the Lyapunov times $T_\Lambda$ extracted from the simulations of Ref.~\onlinecite{Benettin2018FPUT}.  The dashed black line is a fit of   $T_\Lambda$ for $SABA_2$ using linear regression between $\tau=0.013$ and $\tau=0.42$. 
    }
    \label{fig:4}
\end{figure*}

\begin{figure*}
    \centering
    \includegraphics[trim={7.5cm 0.1cm 6cm 0.1cm},clip, scale=0.65]{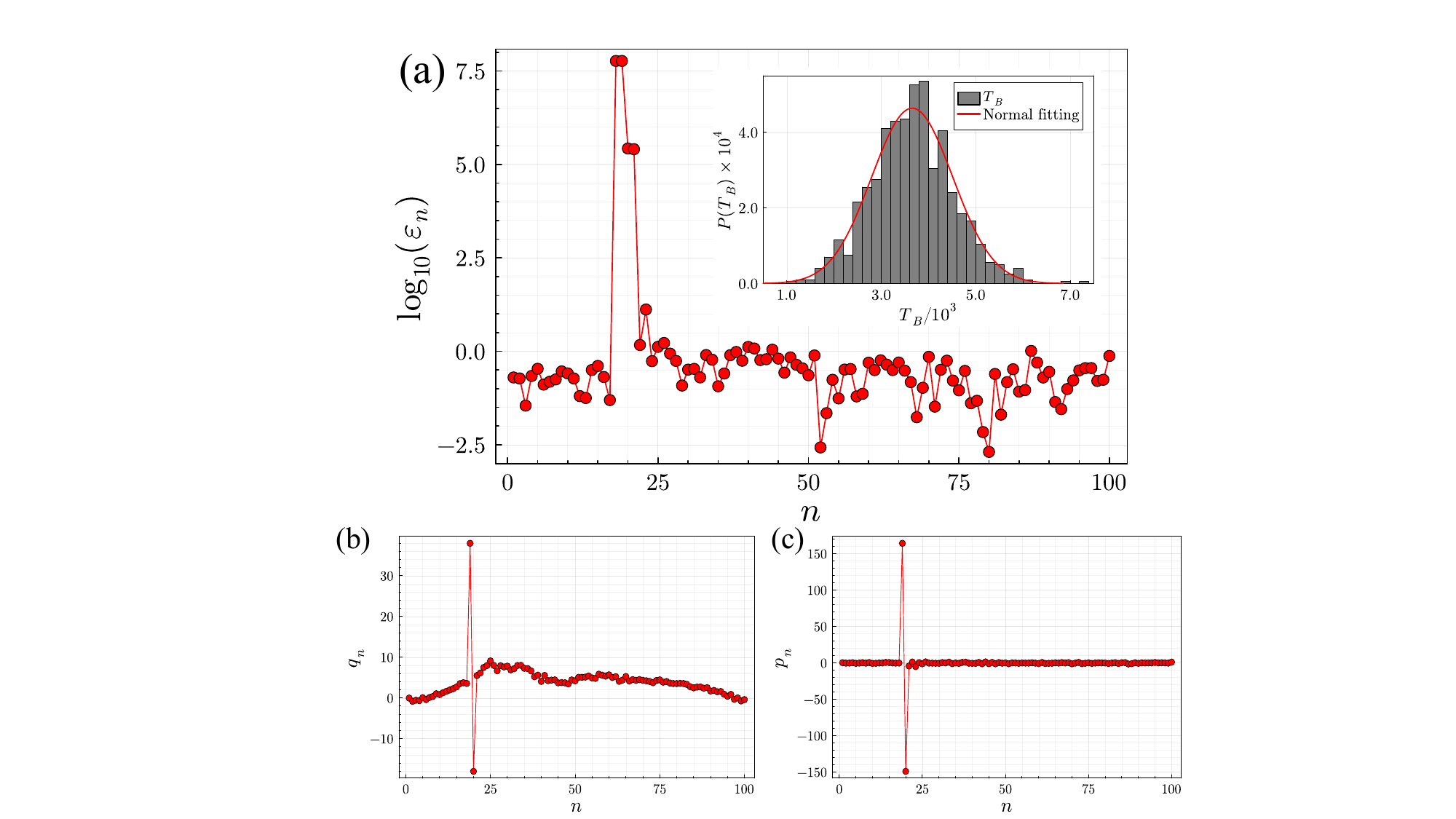}
    \caption{A snapshot of the time-discretized  dynamics of the Toda chain  generated with $SABA2$ symplectic integrator  just before the breakdown.  Panels (a), (b) and (c) show the energy densities (per lattice site), lattice displacements, and lattice momenta, respectively, at $t = 2427$ for time step $\tau = 1.0$ and anharmonicity parameter $\alpha=0.25$.   Notice that displacements and momenta at two neighboring sites ($n=19$ and 20) are large in magnitude and opposite in sign. The inset of panel (a) shows the distribution of the breakdown times $T_B$ for $1000$ random  initial conditions A Gaussian fit of this distribution yields a mean $\mu = 3655$, standard deviation $\sigma = 859$, skewness $0.17$, and kurtosis  $0.35$. We generate the initial conditions for this figure as before (see footnote~\ref{ftnt1} on page~\pageref{ftnt1}) only using the $SABA2$ integrator instead of ABA864.
}
    \label{fig:6}
\end{figure*}

To estimate the Lyapunov time $T_\Lambda$ from the time-evolution of $X_1(t)$ shown in Fig.~\ref{fig:2}(a), we adopt the following protocol:\\
(i) 
find the time $\hat{t}$ when $X_1(t)$ reaches its minimum value (if $X_1(t)$  does not saturate,  we take the last point,   i.e., $\hat{t} = T$, where $T$ is the total simulation time); \\
(ii) 
fix the time window  $I_{\hat{t}} = [0.1 \hat{t} , \hat{t}]$, such that the end points differ by one order of magnitude; \\
(iii)  
define the largest Lyapunov exponent $\Lambda$ as the average of $X_1$ over the interval $I_{\hat{t}}$, i.e.,  $\Lambda =   \langle X_1(t)\rangle_{t\in I_{\hat{t}}} $. \\
 Fig.~\ref{fig:3} displays the Lyapunov time $T_\Lambda$ as a function of the time-step $\tau$ for three difference system sizes: $N=100, 250,$ and $500$  shown with solid black circles, red squares, and green diamonds, respectively. 
We find that $T_\Lambda$ grows monotonously with decreasing $\tau$ covering more than three decades, while $\tau$  varies only over one decade.
The three curves show excellent overlap   revealing that $T_\Lambda$ is essentially independent of the system size $N$.
Our data indicate that $T_\Lambda(\tau\rightarrow 0) \rightarrow \infty$ 
in agreement with the fact that the Hamiltonian becomes integrable in the limit $\tau\rightarrow 0$.  
We also plot the breakdown time $T_B$  in Fig.~\ref{fig:3} using empty symbols (as opposed to solid symbols  for $T_\Lambda$). 
 We report $T_B$   only for $\tau=1$ and $\tau=0.9$. These two values of $T_B$ correspond to the vertical dashed lines   in Fig.~\ref{fig:2}.  Similar to $T_\Lambda$, the breakdown time $T_B$
does not show a strong dependence on the system size.  Note that $T_B$ is at least three orders of magnitudes larger than the corresponding Lyapunov time indicating a clear scale separation.

The  integrator $ABA864$ we used above is  fourth order, $p=4$.
To compare different symplectic integrators and to collect more data for $T_B$, we repeat the computations with a second-order, $p=2$, integrator  $SABA_2$~\cite{skokos2009deloc} for $N=100$ (see Appendix~\ref{sec:A2b} for the description of this integrator). 
We  observe from Fig.~\ref{fig:4}(b) that the values of $T_\Lambda$  provided by the $SABA_2$ scheme (red squares) are smaller than those  provided by $ABA864$   (black dots) by  1-2 orders of magnitude.  This is to be expected, since the $SABA_2$ scheme is less accurate and therefore  the dynamics it generates  is further from its integrable Toda limit enabling a stronger chaos   as compared to the $ABA864$ dynamics with the same $\tau$. 
In addition to $T_\Lambda$ for $SABA_2$ and $ABA864$, we also show in Fig.~\ref{fig:4}(b) the values of $T_\Lambda$ which we extracted from the simulation of Toda dynamics by Benettin \textit{et. al.} who used a different integrator~\cite{Benettin2018FPUT}.
 In all three cases $T_\Lambda$ apparently diverges as $\tau\rightarrow 0$. The divergence appears to be a power-law-like, $T_\Lambda\propto \tau^{-\eta}$ for small $\tau$ with an integrator-dependent exponent $\eta>0$. It is slowest for $SABA_2$  in which case we have data in a sufficiently large range of $\tau$ for a reliable fit. In this case, we find $\eta\approx 1.36$.

Fig.~\ref{fig:4}(a)  shows the time evolution of the finite-time largest Lyapunov exponent $X_1(t)$ for   $SABA_2$. Notice that in this case the integration visibly breaks down at four different values of $\tau$ within the same time interval as that shown in Fig.~\ref{fig:2}. The measured breakdown times $T_B$ are highlighted by the vertical dashed lines.  The ratio $T_B/T_{\Lambda}$ increases from a value of  order 1 at $\tau=1$ to a value of  order  $10^4$ for $\tau=0.6$ clearly indicating that the two time scales $T_{\Lambda}$
and $T_B\gg T_\Lambda$ scale differently with $\tau$ and quickly separate as $\tau$ grows. They must be therefore due to two distinct   features of the discretized Toda chain dynamics. 
Notice also that unlike $T_\Lambda$, the breakdown times $T_B$ for  $SABA_2$ are substantially lower  than those those  for  $ABA864$ [see the vertical dashed lines in Fig.~\ref{fig:2}(a) and (b)].

As mentioned above, in the case of the Toda chain we were able to identify the precise sequence of events (mechanism) that leads to the breakdown.  The two key ingredients of this mechanism are  the split-step nature of symplectic integrators, where the system evolves ballistically (linearly in time) during each time step and the exponential dependence of the Toda potential on the coordinates. Recall that these integrators split the Toda Hamiltonian $H$ into its kinetic ($A$)  and potential ($B$) parts and evolve the system with $A$ over some of the time steps and and with $B$ over the rest. For simplicity, let us consider the simplest integrator where all time intervals are of the same length $\tau$ and suppose  we evolve with $A$ over odd intervals and with $B$ over the even. For the Toda chain this   evolution with $A$ and $B$ is given by Eq.~(\ref{eq:toda_res}) in Appendix~\ref{sec:A1a}. The evolution with $A$ does not change the momenta $p_n$ and changes the coordinates as
\begin{equation}
q'_n=q_n+\tau p_n.
\label{A}
\end{equation}
Similarly, the evolution with $B$ conserves the coordinates but changes the momenta as  
\begin{equation}
p_n' =  p_n + 
   \frac{\tau}{2\alpha}\left[ e^{2\alpha(q_{n+1}-q_n)} - e^{2\alpha(q_{n}-q_{n-1})}  \right].
   \label{B}
 \end{equation}
 
Since this discretization of the time evolution breaks the integrability and since the Hamiltonian is now time dependent, there are presumably no conserved quantities and we expect the system to explore the entire phase space as discussed earlier in this section. Then, eventually, there will be a fluctuation such that the magnitude of at least one of the coordinates or momenta is large. For definiteness, take $q_k$ to be large in magnitude, $\lvert q_k\rvert\gg1$, and negative and assume that
   the magnitudes of the rest of coordinates and momenta, the anharmonicity parameter $\alpha$, and the time step $\tau$ are all of order 1. We also take $k$ to be away from the end points (specifically, $2<k<N$).  Starting in this state and evolving with $B$ over one time step according to Eq.~\eqref{B}, we obtain, up to a prefactor of order 1,
 \beg
 p'_k\approx-p'_{k+1}\approx \frac{\tau}{2\alpha} e^{2\alpha \lvert q_k\rvert },
 \label{ps}
 \en
 where we only kept the exponentially large terms. The rest of coordinates and momenta remain of order 1. Next, we have to evolve with $A$ according to \eref{A}. This results in
 \beg
  q'_k\approx-q'_{k+1}\approx \frac{\tau^2}{2\alpha} e^{2\alpha \lvert q_k\rvert }.
  \label{qs}
  \en
 Subsequent evolution with $B$  produces four extremely large momenta, $p''_{k-1}, p''_k, p''_{k+1},$ and $p''_{k+2}$, approximately equal in magnitude to an exponent of an exponent of a large number   
 \beg
 p''_{k\pm1}\approx p''_{(k+1)\pm1}\approx \frac{\tau}{2\alpha} \exp\left[\tau^2 e^{2\alpha \lvert q_k\rvert } \right]
 \label{expexp}
 \en
 leading to the breakdown of the simulation at this or at most the next $B$-step depending on the value of $\lvert q_k\rvert$.
 
 We check this picture against numerics in Fig.~\ref{fig:6}. The  $SABA2$  symplectic integrator with which this figure is generated is different from the simplest integrator in the above argument. Nevertheless, the main idea is the same. We see from Fig.~\ref{fig:6} that just before the breakdown the displacements (coordinates) and momenta at two neighboring sites $k=18$ and $k+1=19$ are much larger than the rest, roughly equal in magnitude, and opposite in sign. This agrees with Eqs.~\eqref{ps} and \eqref{qs}. Further, these equations give four values of $\lvert q_k\rvert$ that range from about 5 to 9 with an average of about 7.0.   Using this average value in \eref{expexp} together with $\tau=1$ and $\alpha=0.25$, we obtain a number larger than $10^{14}$. The next $B$-step  must produce an exponent of this number, which is much larger than $10^{308}$ -- the largest  number  available in double precision in standard Fortran compilers. At this point the numerical simulation  breaks down. We also show in Fig.~\ref{fig:6} the distribution of the breakdown times $T_B$ for randomly generated initial conditions, which appears to be roughly Gaussian, and the energy density per site just before the breakdown. The latter is defined as
\begin{equation}
 \label{eq:toda_eq_dim_less}
\begin{cases}
\varepsilon_1 = \frac{p_1^2}{2} + \frac{1}{2}V_\mathrm{T}(q_{2}-q_1) + V_\mathrm{T}(q_1),  \\
\varepsilon_n = \frac{p_n^2}{2} + \frac{1}{2}V_\mathrm{T}(q_{n+1}-q_n) + \frac{1}{2}V_\mathrm{T}(q_n -q_{n-1}), \\
\hspace{10em} \text{for } n = 2, \ldots, (N-1), \\
\varepsilon_N = \frac{p_N^2}{2} + V_\mathrm{T}(-q_N) + \frac{1}{2}V_\mathrm{T}(q_N - q_{N-1}).
\end{cases}
\end{equation}

\begin{figure*}
    \centering
    \includegraphics[trim={0.1cm 4cm 0.1cm 3cm},clip, scale=0.5]{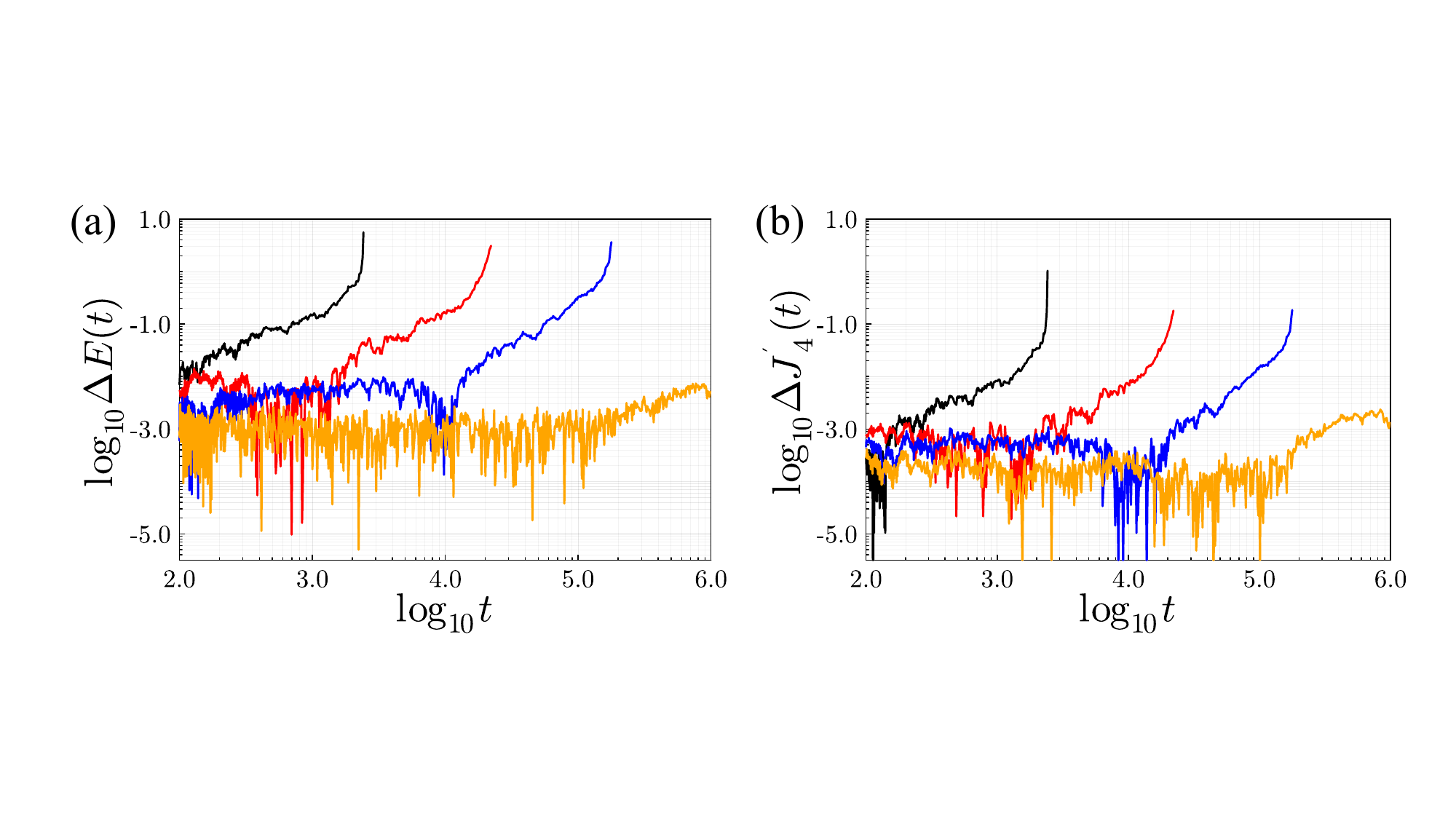}
    \caption{ Relative errors in (a)   energy, $\Delta E(t)$, and (b)   the  nontrivial integral of motion [see Eq.~(\ref{eq:J4Prime})], $\Delta J_4^\prime(t)$,    for the the Toda chain dynamics  generated with  $SABA_2$ split-step symplectic integrator. Different colors correspond to different values of the time-step $\tau$: $\tau=1.0$ (black), $\tau=0.9$ (red), $\tau=0.75$ (blue), and $\tau=0.56$ (orange).  The initial condition is the same as in Fig.~\ref{fig:6}. 
    }
    \label{fig:5}
\end{figure*}

\begin{figure*}
    \centering
    \includegraphics[trim={0.1cm 4cm 0.1cm 3cm},clip, scale=0.5]{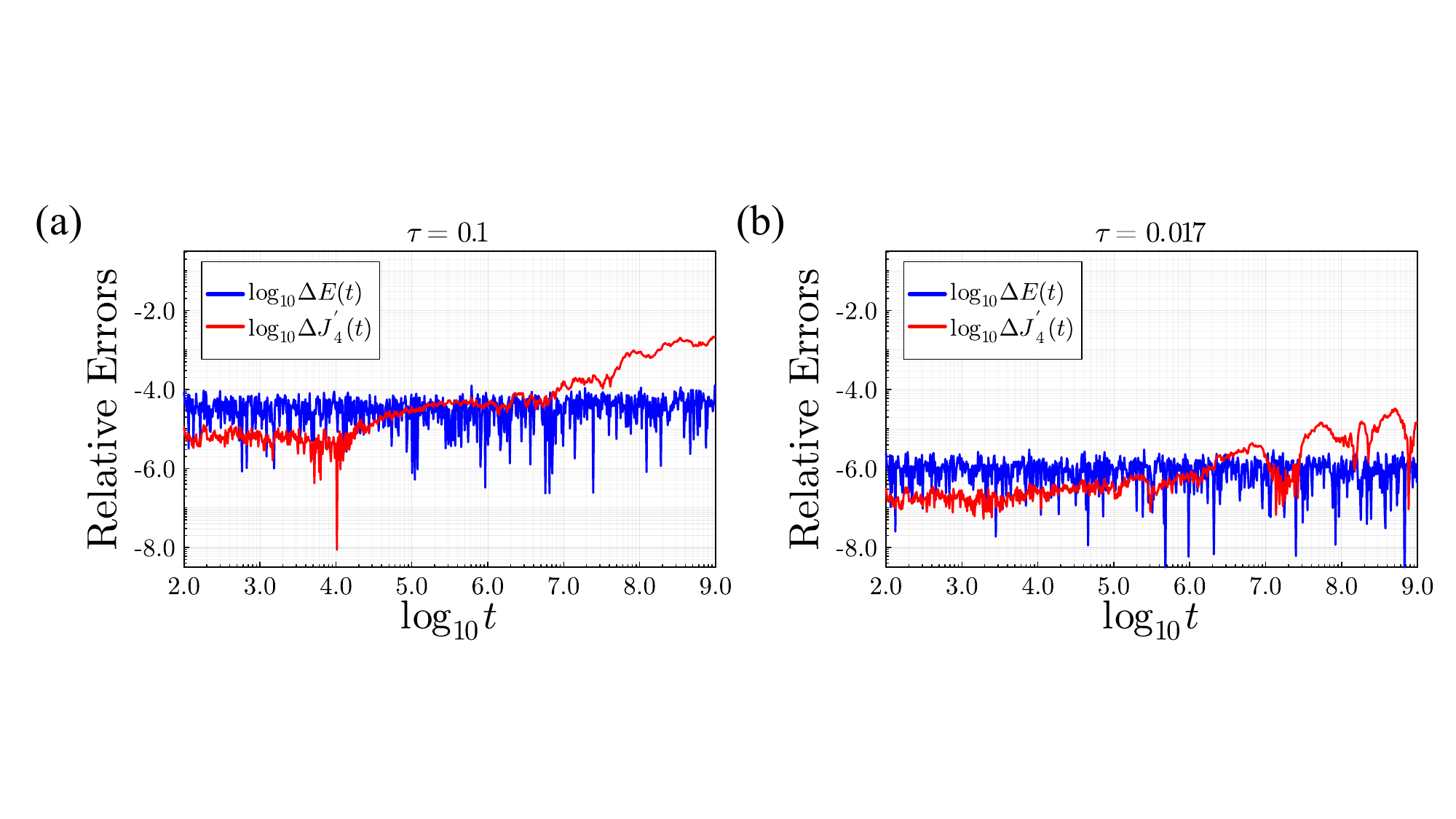}
    \caption{ Comparison of relative errors in energy $\Delta E(t)$ (blue) and the  nontrivial integral of motion $\Delta J_4^\prime(t)$ (red) with small step sizes for the the Toda chain dynamics  generated with the $SABA_2$ split-step symplectic integrator. In panels (a) and (b), we show the relative errors for $\tau = 0.1$ and $\tau = 0.017$ with corresponding Lyapunov times ($T_\Lambda$) being $1.4 \times 10^5$ and $1.4 \times 10^6$, respectively. We estimate $T_J$ by the time corresponding to the crossings of the two relative error plots. In these plots, we indeed have $T_J \approx T_\Lambda$.
    }
    \label{fig:7}
\end{figure*}

 Since the underlying Toda chain is integrable, it is of interest to compare the fluctuations (relative error) in energy and in nontrivial integrals of motion. We do so in Fig.~\ref{fig:5}. The  next integral of motion after the energy in the hierarchy of the integrals of motion for the Toda chain is $J'_4$ given by Eq.~(\ref{eq:J4Prime}) in Appendix~\ref{sec:A4}. We define the relative error in $J'_4$ as in \eref{eq:rel_err} but with the replacement $H\to J'_4$. We see from Fig.~\ref{fig:5} that the relative errors in the energy and the integral of motion behave similarly and run away roughly at the same breakdown times for $1 \geq \tau \geq 0.56$. As a matter of fact, up to a factor of 10, all time scales $T_\Lambda \approx T_J\approx T_E \ll T_B$ coincide. In order to observe the differences between these time scales, we choose smaller values of $\tau$ in Fig.~\ref{fig:7}. We clearly observe that $T_E$ grows much faster with diminishing $\tau$ than $T_\Lambda \approx T_J$. If enough time span is given between $T_J$ and $T_E$, $J$ may start to behave ergodically as in any other nonintegrable Hamiltonian system like the FPUT one.

\section{Conclusion}\label{sec:conclusions}

Here we studied the long-time dynamics of a classical integrable lattice model -- Toda chain with fixed ends  -- with the help of split-step symplectic integrators.  
We made two key observations.  First, time discretization  (more generally, the approximate nature of the integration method) breaks the integrability and induces chaos. We then analyze the properties of such dynamical chaos. We find that the original regular (non-chaotic) dynamics of the Toda chain becomes chaotic with the maximum Lyapunov exponent $\Lambda(\tau)$ controlled by the  time-discretization parameter $\tau$. In the limit $\tau\to0$, $\Lambda(\tau)$ vanishes as a power law. The fact that $\Lambda(\tau)\to0$ as $\tau\to0$ is not surprising as in this limit we should fully recover the integrable Toda dynamics.  In contrast, we saw that for the nonintegrable Fermi-Pasta-Ulam-Tsingou chain $\Lambda(\tau)$, while also initially decreasing with decreasing $\tau$, tends to a finite value when $\tau\to0$. Importantly, chaos manifests itself well before the relative energy error $\Delta E(t)$ (typically used  to track the stability and accuracy of an integrator) becomes significant. Therefore, symplectic schemes applied to integrable systems may result in chaotic dynamics unbeknownst to the reader of the numerical study, hence, rendering the   simulations wrong.

Second, we saw that the energy starts growing and the system starts heating up in a Floquet manner at a time $T_E(\tau)$.  A subsequent breakdown  (dramatic loss of accuracy) of the simulation occurs at a timescale $T_B(\tau) \gg T_E(\tau)$ much larger than the Lyapunov time $T_\Lambda(\tau)=\Lambda^{-1}(\tau)$ at which chaos becomes apparent. The separation of timescales is especially obvious for small $\tau$ as $T_B \gg T_E \gg T_\Lambda$
as $\tau\to0$. 
For times $t < T_\Lambda$ the system evolves as the integrable Toda chain. For $T_\Lambda < t < T_E$ the system evolves as a nonintegrable Hamiltonian perturbation of the Toda chain (e.g. the FPUT one) conserving the energy, being chaotic, and resulting in the loss of conservation of other Toda integrals of motion at times $T_J \approx T_\Lambda$.
For $T_E < t < T_B$ the system enters a Floquet regime, and
heats up towards a featureless infinite temperature state~\cite{PhysRevX.4.041048,PhysRevE.90.012110,PONTE2015196,10.21468/SciPostPhys.3.4.029}. As a result, the system explores the entire phase space. Since the latter is noncompact   for the Toda lattice,  it eventually reaches extremely large values of the coordinates and momenta beyond the ability of the computer to handle. 

Moreover, we were able to pinpoint the specific mechanism  of the breakdown  for the Toda chain.  In this case, it is due to the split-step nature of the integrator, which implies ballistic (linear in time) evolution during each time step, coupled with the exponential dependence of the Toda potential on the coordinates.  We demonstrated using the discretized equations of motion  that as soon as at least one coordinate or momentum becomes large, an irreversible divergence to infinity takes place. 

Our results are very general and applicable to a broad range of integrable classical and quantum interacting many-body models. Suppose, for example, we quantize the Toda chain by promoting coordinates $q_n$ and momenta $p_n$ to quantum operators, such that $[q_n, p_n]=i$. The quantum Toda chain is also integrable~\cite{Olshanetsky:1977aa,GUTZWILLER1981304,10.1007/3-540-15213-X_80,Pasquier:1992aa}, and we anticipate that chaos will again ensue as a result of time discretization (trotterization). Similarly, at a much larger timescale $T_B$ we expect a breakdown of the numerical simulation. Namely, the expectation values of $q_n$ and $p_n$
will grow extremely large, and the location where most of the weight of the many-body wavefunction $\Psi(q_1, q_2,\dots)$ is concentrated will move to infinity.


Interestingly, no breakdown can occur when the phase space of the classical model is compact, or, for a quantum model, the dimensionality of the Hilbert space is finite.  For example, consider classical spin and quantum spin-$\frac{1}{2}$ models. The infinite temperature state is the state where each spin points in a random direction independent of the other spins. 
The numerical simulation  should have no fundamental  difficulty approaching this state as there are no divergencies along the way. 

Our study has  implications for evaluating errors in  quantum simulations in quantum information science. Here the goal is to determine the time evolution of a prescribed Hamiltonian, and one of the main approaches is precisely the splitting method~\cite{suzuki1991general, berry2007efficient, poulin2014trotter, babbush2016exponentially, pitsios2017photonic, tranter2019ordering, cirstoiu2020variational, bolens2021reinforcement, lin2021real, richter2021simulating, tepaske2022optimal, ZhaoBukovHeylMoessner2022} (aka Trotterization in this context) we used in this paper for the classical Toda chain. Our results indicate that Trotterization errors can be complex, i.e., vary significantly between   observables and  qualitatively affect the character   of the dynamics (chaotic vs. regular) when the quantum Hamiltonian which we are attempting to simulate is integrable. In a recent study \cite{UnitaryCktWuCai}, an isolated quantum system, whose time evolution is described by a sequence of unitary maps, was shown to display artificial dissipation induced by time discretization.






\begin{acknowledgments}
We thank R. McLachlan for pointing to relevant literature.
This work was supported by the Institute for Basic Science, Project Code (Project No. IBS-R024-D1).
\end{acknowledgments}










\appendix

\section{Resolvent operators and variational problems for Toda and FPUT}\label{sec:A1}

In this appendix, we detail the equations of motion~\eqref{eq:ham_eq}, variational equations~\eqref{eq:var_eq} and their respective resolvents for both Toda and FPUT models.

\subsection{Toda chain}\label{sec:A1a}

The   Hamiltonian~\eqref{eq:sep_ham} of the Toda chain with fixed boundary condition  
in terms of coordinates $q_n$ and canonically conjugate momenta $p_n$ reads
\begin{multline}
\label{eq:toda_ham}
H_\mathrm{T} = \Bigg[\sum_{n=1}^{N} \frac{p_n^2}{2}\Bigg] + \Bigg[ \frac{e^{2\alpha q_{1}} - 2\alpha q_{1} -1 }{4\alpha^2} \\
+ \sum_{n=1}^{N-1} \frac{e^{2\alpha(q_{n+1}-q_n)} - 2\alpha(q_{n+1}-q_n) -1 }{4\alpha^2} \\
+ \frac{e^{-2\alpha q_{N}} + 2\alpha q_{N} -1 }{4\alpha^2}\Bigg] = A+B
\end{multline}
For the fixed boundary condition ($q_{0} = p_{0} = q_{N+1} = p_{N+1} = 0$),  Hamilton's  equations of motion~\eqref{eq:ham_eq} become 
\begin{equation}
\label{eq:toda_eq}
\begin{cases}
\dot{q}_n = p_n, \quad \text{for } n = 1, \ldots, N, \\
\dot{p}_1 = \frac{1}{2\alpha}\left[ e^{2\alpha(q_{2}-q_1)} - e^{2\alpha q_{1}}  \right], \\ 
\dot{p}_n = \frac{1}{2\alpha}\left[ e^{2\alpha(q_{n+1}-q_n)} - e^{2\alpha(q_{n}-q_{n-1})}  \right],\\
\hspace{10em} \text{for } n = 2, \ldots, (N-1), \\
\dot{p}_N = \frac{1}{2\alpha}\left[ e^{- 2\alpha q_N} - e^{2\alpha(q_{N}-q_{N-1})}  \right].
\end{cases}
\end{equation}
The variational equations~\eqref{eq:var_eq}   for the deviations  $\{ \delta q_n,\delta p_n\}$ take the form 
\begin{equation}
\label{eq:toda_tangeq}
\begin{cases}
\dot{\delta q}_n = \delta p_n, \quad \text{for } n = 1, \ldots, N, \\
\dot{\delta p}_1 = - [ e^{2\alpha(q_{2}-q_1)} + e^{2\alpha q_{n}}  ]\delta  q_1 + [ e^{2\alpha(q_{2}-q_1)} ]\delta  q_{2}, \\
\dot{\delta p}_n = - [ e^{2\alpha(q_{n+1}-q_n)} + e^{2\alpha(q_{n}-q_{n-1})}  ]\delta  q_n  \\ 
\qquad + [ e^{2\alpha(q_{n+1}-q_n)} ]\delta  q_{n+1} 
+ [ e^{2\alpha(q_{n}-q_{n-1})}]\delta  q_{n-1}, \\
\hspace{10em} \text{for } n = 2, \ldots, (N-1), \\
\dot{\delta p}_N = - [ e^{-2\alpha q_N} + e^{2\alpha (q_{N}-q_{N-1})}  ]\delta  q_N \\
+ [ e^{2\alpha(q_{N}-q_{N-1})} ]\delta  q_{N-1}.
\end{cases}
\end{equation}
Splitting the Toda Hamiltonian~\eqref{eq:toda_ham} into the kinetic and potential parts, $H= A+B$, yields the following two systems of differential equations that describe the evolution with $A$ alone ($L_A$) and with $B$ alone ($L_B$)
\begin{subequations}
\begin{equation}
L_{A}\colon \begin{cases}
\dot{q}_n = p_n, \quad \text{for } n = 1, \ldots, N, \\
\dot{p}_n = 0, \quad \text{for } n = 1, \ldots, N, \\
\dot{\delta q}_n = \delta p_n, \quad \text{for } n = 1, \ldots, N, \\
\dot{\delta p}_n = 0, \quad \text{for } n = 1, \ldots, N,
\end{cases}
\end{equation}
\begin{equation}
L_{B}\colon \begin{cases}
\dot{q}_n = 0, \quad \text{for } n = 1, \ldots, N, \\
\dot{p}_1 = \frac{1}{2\alpha}\left[ e^{2\alpha(q_{2}-q_1)} - e^{2\alpha q_{1}}  \right], \\ 
\dot{p}_n = \frac{1}{2\alpha}\left[ e^{2\alpha(q_{n+1}-q_n)} - e^{2\alpha(q_{n}-q_{n-1})}  \right],\\
\hspace{10em} \text{for } n = 2, \ldots, (N-1), \\
\dot{p}_N = \frac{1}{2\alpha}\left[ e^{- 2\alpha q_N} - e^{2\alpha(q_{N}-q_{N-1})}  \right], \\
\dot{\delta q}_n = 0, \quad \text{for } n = 1, \ldots, N, \\
\dot{\delta p}_1 = - [ e^{2\alpha(q_{2}-q_1)} + e^{2\alpha q_{n}}  ]\delta  q_1 + [ e^{2\alpha(q_{2}-q_1)} ]\delta  q_{2}, \\
\dot{\delta p}_n = - [ e^{2\alpha(q_{n+1}-q_n)} + e^{2\alpha(q_{n}-q_{n-1})}  ]\delta  q_n  \\ 
\qquad + [ e^{2\alpha(q_{n+1}-q_n)} ]\delta  q_{n+1} 
+ [ e^{2\alpha(q_{n}-q_{n-1})}]\delta  q_{n-1}, \\
\hspace{10em} \text{for } n = 2, \ldots, (N-1), \\
\dot{\delta p}_N = - [ e^{-2\alpha q_N} + e^{2\alpha (q_{N}-q_{N-1})}  ]\delta  q_N \\
\hspace{10em} + [ e^{2\alpha(q_{N}-q_{N-1})} ]\delta  q_{N-1}.
\end{cases}
\label{eq:toda_split}
\end{equation}
\end{subequations}
For an advancement by one time-step $\tau$, we integrate the coordinates $\{q_n, p_n, \delta q_n,\delta p_n\}$ at time $t$ to $\{q_n', p_n', \delta q_n',\delta p_n'\}$ at time $t+\tau$
\begin{equation}
\begin{split}
e^{ \tau L_{A}} \colon &
\begin{cases}
    q_n' = q_n + \tau p_n, \quad \text{for } n = 1, \ldots, N,        \\
  p_n' =  p_n, \quad \text{for } n = 1, \ldots, N, \\
  \delta q_n' =\delta q_n +   \tau \delta p_n, \quad \text{for } n = 1, \ldots, N,   \\
    \delta p_n' = \delta  p_n, \quad \text{for } n = 1, \ldots, N,
\end{cases}  
\\
e^{ \tau L_{B}}\colon &
\begin{cases}
    q_n' =    q_n, \quad \text{for } n = 1, \ldots, N,         \\
    p_1' =  p_1 + 
   \frac{\tau}{2\alpha}\left[ e^{2\alpha(q_{2}-q_1)} - e^{2\alpha q_{1}}  \right]  \\
   p_n' =  p_n + 
   \frac{\tau}{2\alpha}\left[ e^{2\alpha(q_{n+1}-q_n)} - e^{2\alpha(q_{n}-q_{n-1})}  \right], \\
   \hspace{10em} \text{for } n = 2, \ldots, (N-1),  \\
   p_N' =  p_N + 
   \frac{\tau}{2\alpha}\left[ e^{- 2\alpha q_N} - e^{2\alpha(q_{N}-q_{N-1})}  \right]  \\
    \delta q_n' =  \delta  q_n, \quad \text{for } n = 1, \ldots, N,   \\
    \delta p_1' =  \delta p_1 + \tau \big\{  - [ e^{2\alpha(q_{2}-q_1)} + e^{2\alpha q_{n}}  ]\delta  q_1 \\
   \hspace{10em} + [ e^{2\alpha(q_{2}-q_1)} ]\delta  q_{2} \big\}, \\ 
    \delta p_n' =  \delta p_n + \tau \big\{  - [ e^{2\alpha(q_{n+1}-q_n)} + e^{2\alpha(q_{n}-q_{n-1})}  ]\delta  q_n \\ 
\qquad+ [ e^{2\alpha(q_{n+1}-q_n)} ]\delta  q_{n+1} 
+ [ e^{2\alpha(q_{n}-q_{n-1})}]\delta  q_{n-1}   \big\} , \\ 
\hspace{10em} \text{for } n = 2, \ldots, (N-1), \\
\delta p_N' =  \delta p_N + \tau \big\{  - [ e^{-2\alpha q_N} + e^{2\alpha (q_{N}-q_{N-1})}  ]\delta  q_N  \\
\hspace{10em} + [ e^{2\alpha(q_{N}-q_{N-1})} ]\delta  q_{N-1} \big\}.
\end{cases}
\end{split}
\label{eq:toda_res}
\end{equation}

\subsection{Fermi-Pasta-Ulam-Tsingou chain}\label{sec:A1b}

The  Hamiltonian of the FPUT chain   reads
\begin{multline}
\label{eq:fput_ham}
H_F = \left[ \sum_{n=0}^{N-1}  \frac{p_n^2}{2} \right]  + \Bigg[ \frac{1}{2}q^2_{1}  + \frac{\alpha}{3}q_{1}^3  \\
+ \sum_{n=1}^{N-1}\Bigg\{ \frac{1}{2}(q_{n+1} - q_{n})^2  + \frac{\alpha}{3}(q_{n+1} - q_{n})^3 \Bigg\} \\
+ \frac{1}{2}q^2_{N}  - \frac{\alpha}{3}q_{N}^3 \Bigg] =A+B.
\end{multline}
The corresponding Hamiltons  equations of motion~\eqref{eq:ham_eq} are 
\begin{equation}
\label{eq:fput_eq}
\begin{cases}
\dot{q}_n = p_n, \quad \text{for } n = 1, \ldots, N, \\
\dot{p}_1 =  (q_{2} -2 q_1) +\alpha \left[ (q_{2} - q_1)^2 - q_1^2 \right], \\
\dot{p}_n =  (q_{n+1} + q_{n-1} -2 q_n) \\
\hspace{5em} +\alpha \left[ (q_{n+1} - q_n)^2 - (q_n - q_{n-1})^2 \right],\\
\hspace{10em} \text{for } n = 2, \ldots, (N-1), \\
\dot{p}_N =  (q_{N-1} -2 q_N) +\alpha \left[ q_N^2 - (q_N - q_{N-1})^2 \right].
\end{cases}
\end{equation}
The variational equations~\eqref{eq:var_eq}   for the deviations $\{ \delta q_n,\delta p_n\}$ become
\begin{equation}
\label{eq:fput_tangeq}
\begin{cases}
\dot{\delta q}_n = \delta p_n, \quad \text{for } n = 1, \ldots, N, \\
\dot{\delta p}_1 =  -[ 2\alpha q_{2} + 2 ]\delta  q_1 + [ 1 +2\alpha (q_{2} - q_1) ]\delta  q_{2},\\
\dot{\delta p}_n =  [ 2\alpha  (q_{n-1} - q_{n+1} ) - 2 ]\delta  q_n \\
\quad + [ 1 +2\alpha (q_{n+1} - q_n) ]\delta  q_{n+1} \\
\hspace{5em} + [1+ 2\alpha (q_{n} - q_{n-1}) ]\delta  q_{n-1}, \\ 
\hspace{10em} \text{for } n = 2, \ldots, (N-1), \\
\dot{\delta p}_N =  [ 2\alpha  q_{N-1} - 2 ]\delta  q_N \\
\hspace{5em} + [1+ 2\alpha (q_{N} - q_{N-1}) ]\delta  q_{N-1}.
\end{cases}
\end{equation}
For the same splitting as in Eq.~\eqref{eq:toda_split}, 
the advancement by a time-step $\tau$ reads
\begin{equation}
\begin{split}
e^{ \tau L_{A}} \colon & 
\begin{cases}
    q_n' = q_n + \tau p_n, \quad \text{for } n = 1, \ldots, N,        \\
  p_n' =  p_n, \quad \text{for } n = 1, \ldots, N, \\
  \delta q_n' =\delta q_n +   \tau \delta p_n, \quad \text{for } n = 1, \ldots, N,   \\
    \delta p_n' = \delta  p_n, \quad \text{for } n = 1, \ldots, N,
\end{cases}  
\\
e^{ \tau L_{B}}\colon &
\begin{cases}
    q_n' =    q_n, \quad \text{for } n = 1, \ldots, N,         \\
    p_1' =  p_1 + 
   \tau \big\{   (q_{2} -2 q_1) +\alpha \left[ (q_{2} - q_1)^2 - q_1^2 \right] \big\}, \\
   p_n' =  p_n + 
   \tau \big\{   (q_{n+1} + q_{n-1} -2 q_n) \\
   \hspace{5em} +\alpha \big[ (q_{n+1} - q_n)^2 - (q_n - q_{n-1})^2\big] \big\}, \\
\hspace{10em} \text{for } n = 2, \ldots, (N-1), \\
    p_N' =  p_N + 
   \tau \big\{   (q_{N-1} -2 q_N) \\
   \hspace{5em} +\alpha \left[ q_N^2 - (q_N - q_{N-1})^2 \right] \big\}, \\
    \delta q_n' =  \delta  q_n, \quad \text{for } n = 1, \ldots, N,   \\
    \delta p_1' =  \delta p_1 + \tau \big\{ -[ 2\alpha q_{2} + 2 ]\delta  q_1 \\
    \hspace{5em} + [ 1 +2\alpha (q_{2} - q_1) ]\delta  q_{2}       \big\}, \\
    \delta p_n' =  \delta p_n + \tau \big\{  [ 2\alpha  (q_{n-1} - q_{n+1} ) - 2 ]\delta  q_n \\
    \hspace{5em} + [ 1 +2\alpha (q_{n+1} - q_n) ]\delta  q_{n+1} \\
    \hspace{7.5em} + [1+ 2\alpha (q_{n} - q_{n-1}) ]\delta  q_{n-1}     \big\}, \\
\hspace{10em} \text{for } n = 2, \ldots, (N-1), \\
    \delta p_N' =  \delta p_N + \tau \big\{ [ 2\alpha  q_{N-1} - 2 ]\delta  q_N  \\
    \hspace{5em} + [1+ 2\alpha (q_{N} - q_{N-1}) ]\delta  q_{N-1}  \big\}.
\end{cases}
\end{split}
\label{eq:FPU_res}
\end{equation}

\section{Symplectic integration schemes}

In this appendix, we present the symplectic integration schemes $ABA864$ and $SABA_2$ used in this work. 

\subsection{$ABA864$}\label{sec:A2a}

The symplectic integration scheme $ABA864$  consists of the following product of resolvent operators $e^{\tau L_{A}}$ and $ e^{\tau L_{B}}$ of addends $A$ and $B$, respectively:
\begin{multline}
  ABA864(\tau ) = e^{a_1 \tau L_{A}}e^{b_1 \tau L_{B}}e^{a_2 \tau L_{A}} e^{b_2 \tau L_{B}}  e^{a_3 \tau L_{A}}e^{b_3 \tau L_{B}} e^{a_4 \tau L_{A}} \\ 
  \times e^{b_4 \tau L_{B}}  e^{a_4 \tau L_{A}}e^{b_3 \tau L_{B}} e^{a_3 \tau L_{A}}e^{b_2 \tau L_{B}} e^{a_2 \tau L_{A}}e^{b_1 \tau L_{B}} e^{a_1 \tau L_{A}} 
\label{eq:aba864}
\end{multline}
for a given time-step $\tau$. 
The coefficients $\{a_1,a_2,a_3,a_4,b_1,b_2,b_3,b_4\}$ are
\begin{equation}
\begin{split}
&a_1 = 0.07113343,  
\qquad \qquad \ \ 
b_1 = 0.18308368, \\ 
&a_2 = 0.24115343,  
\qquad \qquad \ \ 
b_2 = 0.31078286, \\ 
&a_3 = 0.52141176,  
\qquad \qquad \ \  
b_3 = -0.02656462, \\ 
&a_4 = -0.33369862,  
\qquad \qquad
b_4 = 0.06539614. \\ 
\end{split}
\label{eq:aba864_coeff}
\end{equation}
Note that the coefficients in Eq.~\eqref{eq:aba864_coeff} are truncated to the eighth decimal place with respect to those reported in Table 3 of Ref.~\onlinecite{BLANES201358}.

\subsection{$SABA_2$}\label{sec:A2b}

The symplectic integration scheme $ABA864$  is described by the following product of resolvent operators $e^{\tau L_{A}}$ and $ e^{\tau L_{B}}$ of addends $A$ and $B$, respectively:
\begin{equation}
\begin{split}
  SABA_2(\tau ) &= e^{a_1 \tau L_{A}}e^{b_1 \tau L_{B}}e^{a_2 \tau L_{A}} e^{b_1 \tau L_{B}}e^{a_1 \tau L_{A}} 
\end{split}
\label{eq:saba2}
\end{equation}
for a given time-step $\tau$. 
The coefficients $\{a_1,a_2,b_1\}$ are
\begin{equation}
\begin{split}
&a_1 = \frac{1}{2}\left(1-\frac{1}{\sqrt{3}} \right), 
 \qquad \ \ 
a_2=\frac{1}{\sqrt{3}},
 \qquad \ \ 
b_1 = \frac{1}{2}. \\ 
\end{split}
\label{eq:saba2_coeff}
\end{equation}

\section{The  nontrivial integral for the Toda chain with fixed boundary}\label{sec:A4}

In this appendix, we provide an explicit expression for the first after the total energy nontrivial integral for the Toda chain with fixed boundary conditions (fixed ends) following the approach of Refs.~\onlinecite{Ford1973integrability, Henon1974integrals, Flaschka1974toda, Toda1975studies}. 

We start by introducing the rescaled positions and momenta as 
\begin{equation}
\label{eq:rescaled_p_q}
P_n = -2 \alpha p_n, \qquad Q_n = -2 \alpha q_n,
\end{equation}
for $n = 1, \ldots, N$. In terms of these new variables, the equations of motion take the following dimensionless form:
\begin{equation}
\label{eq:toda_eq_dim_less_1}
\dot{Q}_n = P_n, \quad \dot{P}_n = X_n - X_{n+1},
\end{equation}
where $X_n = e^{-\left(Q_n - Q_{n-1}\right)}$ obeys 
\begin{equation}
\label{eq:toda_eq_X}
    \dot{X}_n = \left(P_{n-1} - P_{n}\right)X_n. 
\end{equation}

To obtain the integrals of motion for the fixed boundary Toda chain with $N$ lattice sites, one starts with a periodic lattice with $(2N + 2)$ sites, where the lattice indices are given by $n = -N, \ldots, N+1$. We then impose the antisymmetric initial conditions  
\begin{equation}
    \begin{split}
      & Q_{-n} = -Q_{n},\quad P_{-n} = -P_{n}, \\ 
      & Q_{0} = Q_{N+1} = 0,\quad P_{0} = P_{N+1} = 0,
    \end{split}
    \label{eq:antisymm_init}
\end{equation}
on the periodic lattice, where $n = 1, \ldots, N$. Here $Q_n$ and $P_n$ can take arbitrary values for $n = 1, \ldots, N$. Using the above, one derives 
\begin{equation}
    \label{eq:antisymm_init_X}
    X_{-n} = X_{n+1}, 
\end{equation}
for $n = 0, 1, \ldots, N$.

The antisymmetric condition is respected by the equations of motion at latter times. Therefore, the stretch of the periodic lattice between $n = 1$ and $n = N$ corresponds to the fixed boundary Toda chain with $N$ lattice sites. The periodic lattice has $2N+2$ integrals of motion that are denoted as $J_m^\prime$ for $m = 1, \ldots, (2N+2)$, where the prime denotes that the integrals of motion for the periodic lattice are calculated after imposing the antisymmetric initial condition \eqref{eq:antisymm_init}. Because of the special initial conditions, all the $J_m^\prime$ with $m$ odd are identically equal to zero. Moreover, there is one integral of motion, which reduces to a constant that is independent of positions and momenta. As a result, we denote the $N$ integrals of motion for the fixed boundary Toda chain as $J_m^\prime$ with $m = 2, 4, \ldots, 2N$.

Since $J_2^\prime \propto H_\mathrm{T}$, the first nontrivial integral for the fixed boundary Toda chain is given by $J_4^\prime$.  Consider the expression for $J_4$ 
\begin{multline}
    \label{eq:J4}
    J_4 = \sum_{n = -N}^{N+1} \bigg[ \frac{1}{4} P_n^{4} + P_n^{2} \left( X_n + X_{n+1} \right)  \\
    + P_n P_{n+1} X_{n+1} + \frac{1}{2}X_n^{2} + X_n X_{n+1}  \bigg],
\end{multline}
which is valid for the periodic lattice  for any initial condition, see, e.g., Eq. (12) of Sec. 4.5 of Ref.\ \onlinecite{Toda1975studies}. Using Eqs.\ \eqref{eq:toda_eq_dim_less_1} and \eqref{eq:toda_eq_X}, one can check that indeed $\dot{J}_4 = 0$. 

After imposing the antisymmetric condition \eqref{eq:antisymm_init} and using Eq.\ \eqref{eq:antisymm_init_X}, one obtains the following integral of motion for the fixed boundary Toda chain from Eq.\ \eqref{eq:J4}:
\begin{multline}
    \label{eq:J4Prime}
    J_4^\prime = \sum_{n = 1}^{N} \left[ \frac{1}{2} P_n^{4} + 2P_n^{2} \left( X_n + X_{n+1} \right) + 2X_n X_{n+1}  \right] \\ + 2\left( X_1^2 + X_{N+1}^2 \right) + \sum_{n = 2}^{N} X_n^2 +  2 \sum_{n = 1}^{N-1} P_n P_{n+1} X_{n+1},
\end{multline}
where 
\begin{equation}
    \label{eq:specialX}
    X_1 = e^{-Q_1}, \qquad X_{N+1} = e^{Q_{N}}.  
\end{equation}
Using Eqs.\ \eqref{eq:toda_eq_dim_less_1} and \eqref{eq:toda_eq_X}, we have checked that $\dot{J}^\prime_4 = 0$. The relative error of this first nontrivial integral of motion is defined as
\begin{equation}
\label{eq:rel_err_J4Prime}
\Delta J_4^\prime(t) = 
\bigg\lvert \frac{J_4^\prime(t) - J_4^\prime(0)}{J_4^\prime(0)} \bigg\rvert,
\end{equation} 
which is then plotted in Fig.\ \ref{fig:5}.


\bibliography{sn-bibliography,sergejflach,Aniket}


\end{document}